# Enhancement of the optical gain in GaAs nanocylinders for nanophotonic applications

Jinal Tapar, Saurabh Kishen, Kumar Prashant, Kaushik Nayak, and Naresh Kumar Emani[a]

*Department of Electrical Engineering, Indian Institute of Technology, Hyderabad, 502285, India*

Semiconductor nanolasers based on microdisks, photonic crystal cavities and metallo-dielectric nanocavities have been studied during the last decade for on-chip light source applications. However, practical realization of low threshold, room temperature semiconductor nanolasers is still a challenge due to the large surface-to-volume ratio of the nanostructures, which results in low optical gain, and hence higher lasing threshold. Further, the gain in nanostructures is an important parameter for designing all-dielectric metamaterial based active applications. Here we investigate the impact of p-type doping, compressive strain, and surface recombination on the gain spectrum and the spatial distribution of carriers in GaAs nanocylinders. Our analysis reveals that the lasing threshold can be lowered by choosing the right doping concentration in the active III-V material combined with compressive strain. This combination of strain and p-type doping shows 100x improvement in gain and ~5 times increase in modulation bandwidth for high-speed operation.

**THE MANUSCRIPT**

The current computing and information technology revolution was driven by tremendous advances in semiconductor electronic devices, which are fast approaching fundamental physical limitations. This prompted researchers worldwide to look for radical technologies such as on-chip optical[1] and plasmonic interconnects[2-4] to supplement current high-speed electronics, or even go beyond THz limit of current electronic devices by deploying all-photonic integrated circuits (PICs)[5]. Even though plasmonic devices have an excellent ability to confine light in deep sub-wavelength nanostructures, the inherent dissipative loss due to constituent metal inhibits their practical realization. Alternative plasmonic materials[6] with low losses like heavily doped semiconductors, transparent conducting oxides *etc.* show a promising path to practical plasmonic applications. More recently, the high-refractive-index dielectric and semiconductor

---

[a] Author to whom correspondence should be addressed. naresh@ee.iith.ac.in

Mie nanoresonators[7-10] have drawn significant attention because of their low dissipative loss at visible and near IR wavelengths, and an excellent ability to manipulate light at the nanoscale. The existence of strong electric and magnetic dipole resonances[11] in the dielectric Mie resonators has facilitated a wide range of nanophotonic applications like directional scattering[12-16], metasurfaces[17-19] based wave front manipulation and switching, nonlinear harmonic generation[20], *etc*. Direct bandgap III-V semiconductors based resonant nanostructures and metasurfaces[21] were recently employed in a range of active optoelectronic devices like on-chip light emitters, single photon sources[22] for quantum PICs[23], second harmonic generation (SHG)[24,25], *etc.* because of their strong nonlinearity and smaller free carrier lifetime compared to indirect bandgap materials. Over last decade, there have been a number of nano-scale semiconductor lasers under investigation: photonic crystal lasers[26], plasmonic lasers[27-29], and hybrid photonic-plasmonic mode lasers[30,31]. More recently, directional lasing from GaAs nanoantenna array[32] has been reported using a novel design that takes advantage of both dielectric nanoantenna resonances and bound states in continuum (BIC)[33,34] confinement. All of these recent nanolaser experiments show universally high threshold and most of them require large optical pumping at low temperature.

In the above active applications, the optical gain achievable by the constituent III-V material plays a crucial role. As the size of lasing cavity is shrunk the non-radiative losses become dominant due to larger surface recombination (higher surface-to-volume ratio), and hence cause an increase in the lasing threshold beyond the damage threshold of the nanostructure. This effect is significantly worsened in the case of electrical injection due to high electrical resistance of nanostructures[35]. These constraints motivated us to address the critical need for careful modeling of gain material so as to develop strategies to improve the gain in nanostructures, and thereby realize nanolasers and other active metamaterial-based applications with lower thresholds.

In this work, we demonstrate the quantitative impact of p-type doping, uniaxial and biaxial compressive strain, and surface recombination which influence the gain in nano-structured active semiconductor material. This study gives insights into approaches for improving gain in semiconductor nanostructures, instrumental in realizing a plethora of metamaterial-based active applications at lower threshold. We confine our analysis to bulk GaAs which is one of the most common III-V semiconductors. Although quantum wells (QW) and quantum dots (QD) are known to have gain in range of few 1000-10000 $cm^{-1}$, their lower modal gain restricts their application in nanolasers. Further, in quantum confined structures, the transition matrix element is dependent on polarization of incoming light which adds to the numerical complexity of the analysis[36]. We also note that our analysis can be extended to other III-V semiconductors as well, but with appropriate changes to electronic and optical material parameters.

In order to calculate the effect of aforementioned factors on the gain in GaAs nanostructures, the study is carried out in two steps – (1) Evaluating carrier distribution in nanostructure geometry as a function of pumped energy, and (2) Estimating the realistic gain achievable in nanostructures using full band-structure parameters obtained through density functional theory (DFT) calculations. A full wave 3D simulation is carried out using a commercial finite element method (FEM) software package (COMSOL Multiphysics™, semiconductor and wave optics module). In the simulations, a plane wave of power, $P_{in}$, is incident on a GaAs nanocylinder, which results in generation of excess electron-hole pairs (EHP) required for population inversion. A uniform isotropic generation rate is considered in our analysis since we are mainly interested in bulk semiconductor nanostructures. However, if the nanostructures were fabricated from semiconductor quantum wells then the polarization of incoming light will play a critical role in the numerical analysis of gain. The spatial distribution of the carrier density is modeled using a rate equation, and the effect of surface recombination is introduced by considering surface recombination velocity (SRV) as a simulation parameter (see SI for implementation details). It is a well-known fact that Auger recombination becomes a dominant channel for energy relaxation as the geometrical size of semiconductor nanostructure decreases[37]. Because of localization effects in quantum-well and quantum-dot based optical devices, total momentum need not be conserved, which makes Auger mechanism threshold less[38] and eventually responsible for low quantum yield. The dimensions of our simulated nanostructures are greater than the excitonic Bohr radius of GaAs, *i.e* 13.5 *nm*. In addition, as the simulations are carried out for room temperature applications, thermal energy is way above the confinement energy for considered dimensions of our nanostructure. Hence, in our simulations, we have incorporated the Auger process having cubic dependence on carrier density as in bulk material based on experimentally determined Auger Coefficient *C*. We have taken the values of Auger coefficient to be $C_n = 2.8 \times 10^{-30}$ cm$^6$/s, $C_p = 9.9 \times 10^{-31}$ cm$^6$/s from literature[39,40] throughout the simulation study. To estimate the realistic gain achievable with obtained carrier density in given GaAs nanocylinder, other parameters like joint density of states ($\rho_r$), quasi-Fermi levels ($E_{fc}$ and $E_{fv}$), Fermi factor ($f_2 - f_1$) *etc.* were calculated based on GaAs full band structure DFT analysis. This was carried out using Synopsys® atomistic simulation toolkit QuantumATK (see SI for atomistic modeling details). The band structure of GaAs, with uniaxial and biaxial compressive strain ranging from 0.5% to 2%, has been simulated in QuantumATK using geometry optimization analysis tool by applying appropriate longitudinal stress along specified axes, incorporating crystal orientation influence on gain. Considering the strain relaxation during nano-patterning of strained semiconductor films, we have restricted the computation of band structure with strain values to 2% only even though theoretically, maximum 4% of strain can be incorporated. (See SI for the strain effects like band-edge shifting and band warping on band structure and its impact on density of states (DOS) effective masses). Since the population inversion needed for lasing is degenerate, we used the exact Fermi-Dirac integral[41] for accurate calculation of carrier density and associated quasi-Fermi levels. The electron-electron interaction in this degenerate regime leads to a very small change in bandgap of ~ few *meV* following n$^{1/6}$ dependence with carrier density and hence can be neglected.

Figure 1(a) below illustrates the computed band structure of GaAs along [100] and [111] directions with the curvatures of conduction band (CB), heavy hole (HH), light hole (LH) and split-off (SO) valence band representing their respective density of states (DOS) effective mass. Figure 1(b) depicts the asymmetry in CB and VB curvatures, reflected in terms of DOS available in each band respectively. The DOS for VB increases rapidly from the band edge due to heavier hole effective masses $m_{dh}^* \sim 0.47 m_0$ whereas the higher curvature of CB *i.e* smaller effective mass $m_{de}^* \sim 0.067 m_0$ leads to lesser DOS in CB, almost two orders less than in VB near the band edge. Because of this asymmetry in DOS effective masses, for the same number of injected electron-hole pairs (EHP), $10^{19}$ cm$^{-3}$, the conduction band quasi-Fermi level $E_{fc}$ moves $\left( E_{fc}/E_{fv} = 0.79/0.68 \right) \sim 1.2$ times more than valence band quasi-Fermi level $E_{fv}$ from the equilibrium Fermi level $E_F$. This asymmetry of DOS has serious implications on the gain that can be achieved in a semiconductor under given optical pumping, which is discussed in the next section.

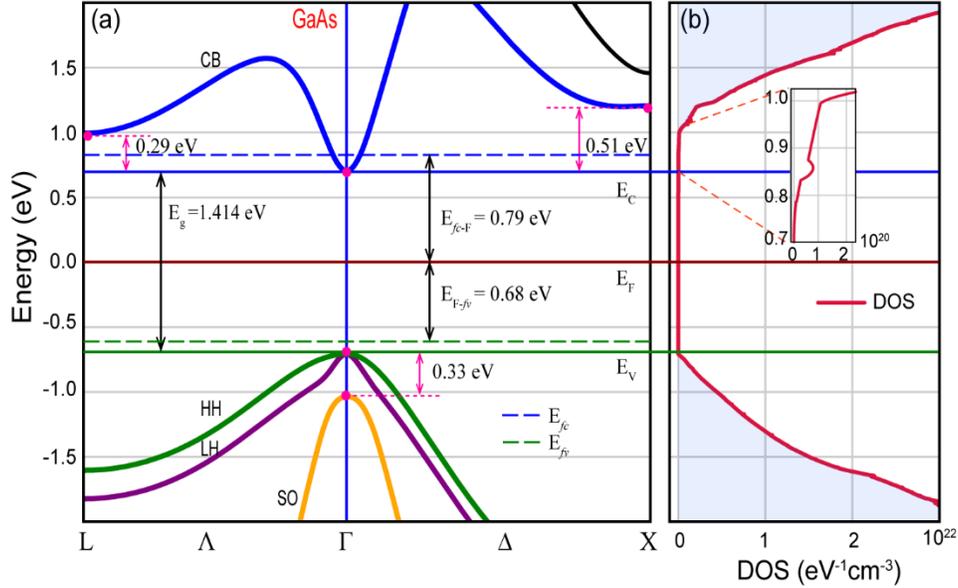

FIG. 1. (a) Energy band diagram of GaAs showing conduction band (CB), heavy hole (HH), light hole (LH) and split-off (SO) valence bands. The curvature of the bands reflects the conduction band and valence band DOS effective mass for GaAs. For $10^{19}$ cm$^{-3}$ EHP generation, CB quasi-Fermi level $E_{fc}$ is inside the CB, but VB quasi-Fermi level $E_{fv}$ is still in bandgap region.

(b) DOS for VB starts increasing rapidly from the VB edge whereas DOS for CB is quite low near to band edge, almost two orders less than VB DOS as shown in zoomed inset.

The equations for modeling gain[42] in our analysis are:

$$g_{21} = g_{max}(E_{hv}) \cdot (f_2 - f_1)$$
$$g_{max}(E_{hv}) = \frac{\pi q^2 \hbar}{n\varepsilon_0 c m_0^2} \frac{1}{hv} |M_T|^2 \rho_r(E_{hv}) \quad (1)$$

where, $g_{21}$ is the gain achievable for the stimulated downward transition $(2 \rightarrow 1)$ and $g_{max}$ is the maximum theoretical gain that can be achieved under complete population inversion. $g_{max}$ is a material property derived from Fermi's Golden rule and is dependent on transition matrix element $|M_T|^2$ and reduced density of states $\rho_r(E_{hv})$. In most scenarios these parameters can be considered as constants for a given material. The Fermi factor $(f_2 - f_1)$ is the most important which determines the gain in lasers. The Fermi occupation probabilities are evaluated using the individual electron and hole energies $E_2$ and $E_1$ in terms of transition energy $E_{hv}$ and the quasi-Fermi energy levels $E_{fc}$ and $E_{fv}$ on high injection of carriers as shown below:

$$f_1 = \frac{1}{e^{(E_1 - E_{fv})/kT} + 1} \quad and \quad f_2 = \frac{1}{e^{(E_2 - E_{fc})/kT} + 1} \quad (2)$$

The net stimulated emission rate, and hence optical gain, becomes positive only when the quasi-Fermi level separation is greater than the photon energy of interest due to degenerate electron and hole ensemble in the active region,

$$E_{fc} - E_{fv} > hv > E_g \quad (3)$$

This condition implies that there should be an excess of electrons and hole concentrations respectively in higher and lower lasing energy levels.

Combining the results obtained from FEM and atomistic simulations with above gain equations, we have determined the carrier density, reduced density of states, quasi-Fermi levels and calculated the gain for intrinsic (undoped, unstrained), only p-doped and, biaxial and uniaxial strained GaAs nanocylinder. The amount and type of strain incorporated in the semiconductor structure modifies the band-structure, the corresponding changes in DOS effective masses and joint DOS are carefully considered while calculating gain for strained structures. The variation of band structure that can be interpreted through calculated Fermi occupation probabilities and gain in each of the cases considered is illustrated in Fig. 2.

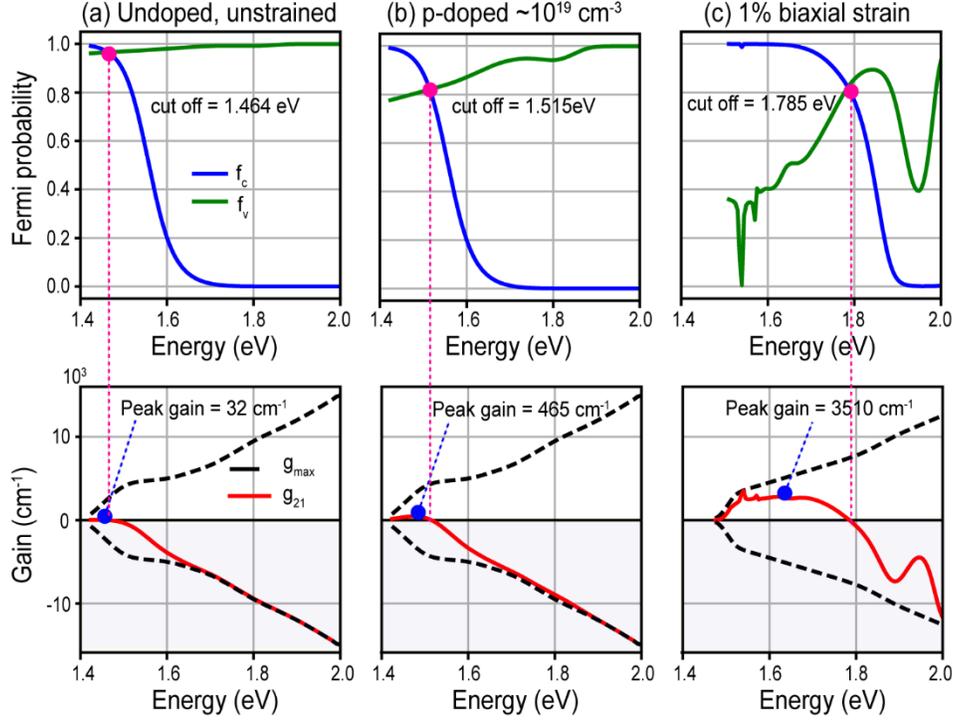

FIG. 2. Fermi probability, gain spectrum (a) For intrinsic GaAs, population inversion is achieved as $(f_2 - f_1)$; negligible gain is observed as net value of Fermi factor $(f_2 - f_1)$ is negligibly small (b) For p-doped GaAs, quasi-Fermi level $E_{fv}$ moves nearer to valence band and 10x improvement in gain is observed as $(f_2 - f_1)$ becomes positive (c) In strained, undoped GaAs, the curvature of valence band is increased, reducing the heavy hole DOS effective mass. The net value of Fermi factor $(f_2 - f_1)$ is considerably larger and ~100x improvement in gain is observed. Along with gain, bandwidth also increases from (a) to (c)

For a given quasi Fermi level separation, there are significant differences in amplitude and bandwidth of the gain. In case of intrinsic GaAs, negligible gain is observed over the spectrum even though the inversion is attained as $(f_2 > f_1)$. The separation of the equilibrium Fermi level into two quasi-Fermi levels $E_{fc}$ and $E_{fv}$, under high pumping conditions, is asymmetrical due to the difference in DOS effective mass of carriers in conduction band $(m^*_{de})$ and valance band $(m^*_{dh})$, as explained in Fig. 1(a). Due to this there are significantly lower number of holes available for recombination. We should note that even though equal number of electrons and holes are introduced due to optical pumping the higher density of states in valence band accommodates more holes per unit energy keeping valence band non-degenerate. On the other hand, conduction band is in the degenerate regime because of the lower DOS available. In order to obtain net optical gain, it becomes necessary to use higher pump intensity creating more EHPs that eventually increases the threshold. This limitation of an intrinsic material can be overcome by using a p-doped semiconductor which is illustrated in Fig. 2(b). We clearly see a small positive difference in the Fermi functions $(f_2 - f_1)$ which allows for more EHP recombination, thereby improving the net gain in the material. Interestingly, Burgess[43] *et al* report increased photoluminescence (PL) emission on zinc (p-type) doping in GaAs nanowires, which is consistent with our analysis that p-type doping in the active material shows significant improvement in the gain.

The incorporation of strain of compressive type in GaAs increases the valence band curvature, greatly reducing the effective mass. In Fig. 2(c), we have shown the case for 1% biaxial compressively-strained, undoped GaAs structure. The reduction in effective mass of holes $(m^*_{dh})$ reduces the density of states in the valence band and causes the quasi-Fermi levels to separate out more symmetrically and hence 100x improvement in gain. Also, the bandwidth (BW = $E_{cutoff} - E_g$, the energy range where the gain is positive) is improved in case of p-doped and strained structure as compared to intrinsic GaAs. The incorporation of p-doping with compressive strain would further increase the gain (not included here). It would be interesting to see how the gain and bandwidth evolves in GaAs nanostructure with varying uniaxial and biaxial compressive strains. Figure 3 shows the direction of applied stress with respect to crystallographic axes aligned to Cartesian axes and compares the peak gain, bandwidth and carrier density for varying amount of uniaxial and biaxial compressive strain in GaAs nanocylinder. All the values for peak gain, bandwidth and carrier density are normalized with the corresponding values of unstrained GaAs to clearly manifest the improvement over unstrained GaAs. In both the cases, peak gain achievable in the nanostructure and bandwidth increases linearly with increasing amount of strain. Gain achieved in uniaxial strained structure is more than in biaxial compressed one. Similarly, the bandwidth is also more for uniaxial strain providing more flexibility in terms of optimum cavity design for tuning spectral overlap of cavity resonance with the material gain spectrum. At the same time, the carrier density needed to achieve the similar gain seems to be minimum for the case of biaxial compressive strain. Depending on the geometry, threshold specifications of a nanolaser design and other fabrication aspects, an optimum choice of type and amount of strain can be engineered for gain enhancement.

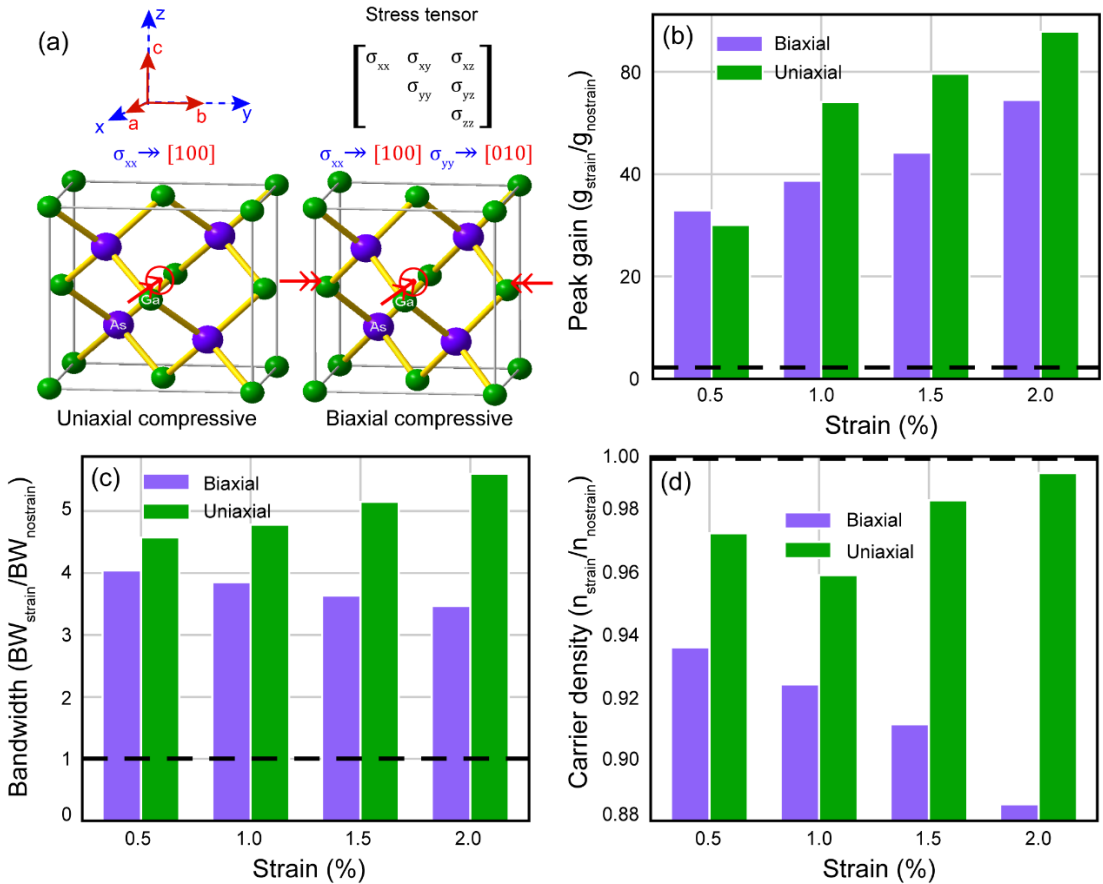

FIG. 3. (a) Direction of applied uniaxial and biaxial compressive stress with respect to crystallographic axes aligned to Cartesian coordinates and elements of stress tensor shown in terms of Miller indices. (b) Peak gain, (c) bandwidth, and (d) carrier density plotted for uniaxial and biaxial compressive strains ranging from 0.5 to 2%, normalized with values for unstrained GaAs; the thick dashes lines represent the unstrained $g_{nostrain} = 71$ $cm^{-1}$, $BW_{nostrain} = 63$ $meV$, $n_{nostrain} = 1.14 \times 10^{20}$ $cm^{-3}$ in (b), (c) and (d)

respectively. For both uniaxial and biaxial cases, peak gain and bandwidth achievable in the nanostructure increases linearly with an increasing amount of strain. On the contrary, the amount of carrier density needed to achieve a similar gain is lesser for biaxial strain than uniaxial, indicating a tradeoff to careful strain engineering.

The observed trend in improvement of gain and bandwidth in uniaxial and biaxial strained nanostructures can be explained through strain modified band structure analysis. With increasing amount of strain in the crystal, the lattice constants in the direction of applied stress gets shortened in case of compressive strain leading to destruction of crystal symmetry. This lowered crystal symmetry results into the following: (i) under biaxial compressive strain, the degeneracy of HH-LH band is lifted and the HH, LH band shifts down and CB band shifts up causing the increase in bandgap, the band curvature of HH band slightly increases and (ii) under uniaxial compressive strain, in addition to degeneracy lifting and CB band shifting up, the mixing between HH and LH bands makes HH and LH bands shift in opposite direction, resulting in a minimal change in the bandgap. The mixing also leads to strong band warping of HH band drastically increasing its curvature and behaves like a LH band near the top of VB (See SI for detailed description of band structure modification under different types of strain). Ideally, maximum gain could have been attained if VB would have achieved the similar band curvature as of CB. The strong band warping in case of uniaxial strain compared to biaxial one leads to more symmetrical separation of quasi-Fermi levels under high EHP injection, and thereby more gain. The reason for improved bandwidth in case of uniaxial strain can be understood from bandgap changes. The cutoff energy $E_{cutoff}$ remains almost same for given amount of both uniaxial and biaxial strain, but the overall increase in bandgap $E_g$ in case of biaxial strain shrinks the bandwidth. We note that even though compressive strain has strong impact on the gain as predicted by Yablonovitch and Kane[44], its practical implementation in terms of fabrication may pose some technological challenges. For example, compressive strain routinely used in semiconductor devices has wafer orientation dependence and when the semiconductor is nano-patterned, there may be geometry dependent strain relaxation.

After discussing about the improvement in gain via p-type doping and strain, we show the impact of surface recombination on the gain in semiconductor nanostructures. Fig. 4(a) illustrates the spatial distribution of injected carriers in an unpassivated GaAs nanocylinder of diameter 350 nm and height 500 nm under optical pumping. Due to significant surface recombination, the injected carrier density is lower in the nanocylinder part compared to the substrate. In smaller volume active region like in VCSEL or nanolasers, the trapping and recombination of injected carriers mediated by dangling bond induced surface states at the surface, depletes the carriers from the bulk. This reduces the internal quantum efficiency of the structure as a substantial part of injected EHPs recombine non-radiatively. Further, due to fewer carriers available for conduction, the electrical resistance of the structure increases making it harder to realize the electrically pumped nanolasers. Thus, surface recombination presents significant hurdles in realizing the optimum performance out of nanolasers, and it cannot be ignored while designing nano-structured lasers.

All the technologically relevant III-V optoelectronic semiconductor materials have high surface recombination velocity of $\sim 10^3$-$10^6$ *cm/s*, GaAs being the worst with highest SRV amongst all with Arsenic atoms contributing towards the surface states. We studied the variation in gain, with uniform pumping power of $P_{in} = 100 \mu W$ and keeping other parameters unchanged. Figure 4 shows the significant reduction of gain with the increasing SRVs for both (b) unstrained and (c) 1% biaxial compressively strained case. Although the maximum gain, $g_{max}$ is less in case of strained material compared to unstrained p-doped material, the overall net gain $g_{21}$ is more under the same pumping conditions. The unstrained material has a higher $g_{max}$ due to the larger density of states in the valence band available to participate in transitions. However, a higher carrier density is necessary to achieve near complete inversion in the unstrained material.

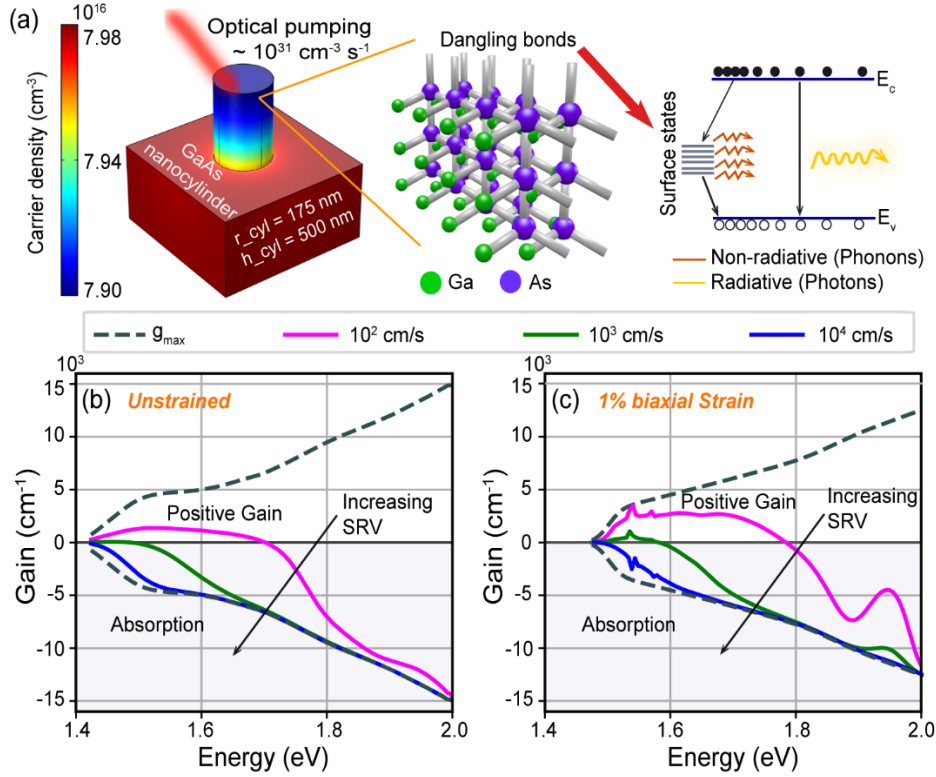

FIG 4. (a) Schematic illustration of the variation in carrier density within the GaAs nanocylinder under optical pumping. The nanocylinder part exhibits lower carrier density when compared to substrate due to dangling bonds at the surface that give rise to recombination via surface states. Gain curves for varied surface recombination velocity (b) Unstrained (c) 1% biaxial compressively strained, GaAs nanostructure. Dashed line denotes maximum gain achievable under complete inversion while the solid lines denote gain for given pumping level. The maximum gain $g_{max}$ is more in unstrained case than strained one due to the larger density of states in the valence band available to participate in transitions. With increasing SRV, gain decreases and all the gain curves transit from gain to absorption as the condition for positive gain as in Eq. 3 is no longer satisfied.

Surface passivation is a well-known approach to overcome the detrimental effects of surface recombination discussed above. Sulfide passivation which has been used for a long time in GaAs devices has poor stability due to atmospheric degradation. Recently, nitride based passivation was proposed in literature[45,46] and it appears to be a promising approach. In principle a higher pumping rate could be used to realize lasing in materials with large SRV without passivation. However, Auger recombination becomes dominant at high levels of carrier injection imposing a higher limit on maximum achievable gain. The strong dependence of Auger process on carrier density and temperature has a detrimental impact on the gain needed for lasing. In essence, harder pumping would not be a good idea and surface passivation or gain enhancement via p-doping or strain is inevitable for nanolaser design. Although for given material, structure and processing steps, SRV is constant, our analysis with varying SRV clearly shows the achievable gain for different levels of surface passivation.

For efficient on-chip optical communication nanolasers must be capable of high-speed operation over larger bandwidth. Higher differential gain *dg/dN* can ideally improve the modulation response of the nanolasers. The differential gain *dg/dN* is a measure of how swiftly the output photon density changes with the change in carrier density. To get a better picture as how the gain varies with carrier density, *i.e.*, input pumped power, the peak of the gain spectrum for varying input power is plotted as a function of sheet carrier density in Fig. 5(a). The p-doped and strained nanostructure has lower transparency carrier density $N_{tr}$ compared to unstrained one. With increased amount of p-doping, transparency carrier density $N_{tr}$ shifts to lower values. But the gain increases at a faster rate for strained case compared to p-doped with the increasing carrier density especially at the band edge due to lesser density of states. Thus, higher differential gain can be attained by having quasi-Fermi levels, aligned with the band edge. Figure 5(b) plots the peak differential gain *dg/dN* for the gain curves in Fig. 5(a). It shows a peak in the graphs for all intrinsic, p-doped and strained closer to transparency carrier density indicating that the nano-laser in this regime is important for high-speed applications. It can be inferred that p-doping is beneficial to lower the transparency carrier density with not much improvement in differential gain while strain improves both transparency condition as well as differential gain.

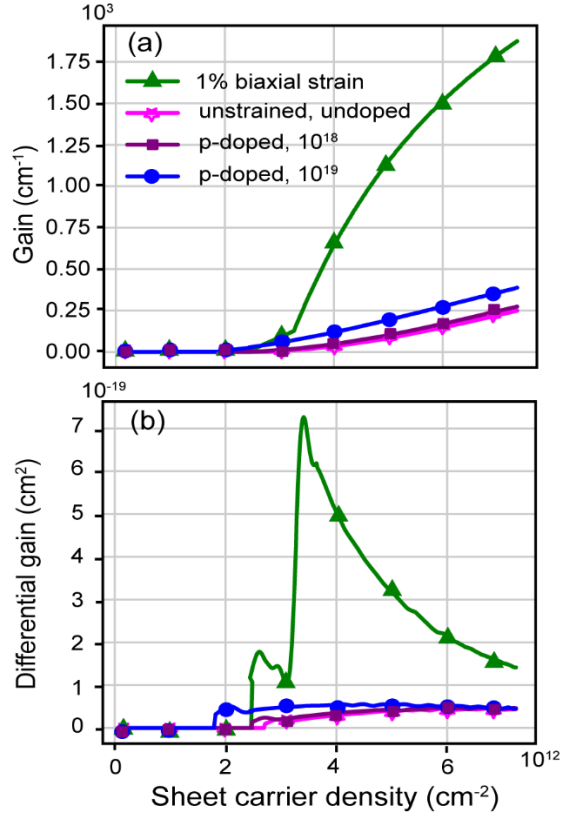

FIG 5. (a) Peak gain versus sheet carrier density in GaAs unstrained-intrinsic, p-doped and strained nanocylinder. The threshold sheet carrier density is lesser and the peak gain value is more for given carrier density in p-doped, and strained GaAs as compared to unstrained-intrinsic (b) Peak differential gain, *dg/dN*, versus sheet carrier density in GaAs unstrained-intrinsic, p-doped, and strained nanostructure. The differential gain value increases faster in response to changing carrier density and so higher modulation speeds can be achieved in strained material.

In the context of development of semiconductor light emitting diodes, Yablonovitch and others have studied the impact of surface recombination in in the microscale QW structures[47]. In recent times, room temperature lasing from surface passivated GaAs-AlGaAs core-shell nanowires[37,48] have been reported. The mitigation of surface recombination issue by surface passivation through AlGaAs cladding can be considered as the main reason for achieving room temperature lasing in this nanowire geometry. To support the need of surface passivation for smaller structures as reported in above literature, we investigate the changes in the gain spectrum and the corresponding gain threshold as the semiconductor nanostructure is scaled into the nanoscale dimensions for aspect ratio ($h_{cyl}/r_{cyl} = 3$) locked nanocylinder with radii in the range 225 nm to 25 nm. We have ignored geometric quantum confinement effects on E(k) dispersion for these nanostructures as their radii are much greater than the excitonic Bohr radius for GaAs, *i.e* 13.5 *nm* at room temperature. The pumping condition and other material parameters were kept same. Clearly, as the radius of nanocylinder reduces, surface area-to-volume *(A/V)* ratio goes up and hence increases the lasing threshold due to non-radiative surface recombination. This can be attributed to the inverse dependence of area-to-volume ratio on the radius of the nanocylinder as in (Eq. 4).

$$R_{surface} = v_s \cdot \frac{A}{V} \cdot N \qquad \frac{A}{V} = \frac{\pi r^2 + 6\pi r^2}{(4/3 \pi r^2 (3r))} = \frac{7}{4r} \qquad (4)$$

Figure 6 plots the peak gain with respect to reciprocal of radius *r*. It clearly depicts the influence of size of nanostructure in deciding the gain threshold for lasing. Even with higher fluence, gain cannot be obtained for smaller volumetric structures because of the lower damage threshold for smaller structures. Inset in Fig. 6 depicts that the bandwidth also decreases with the shrinking cylindrical size, reducing the nanocavity design space for spectral tuning of modes.

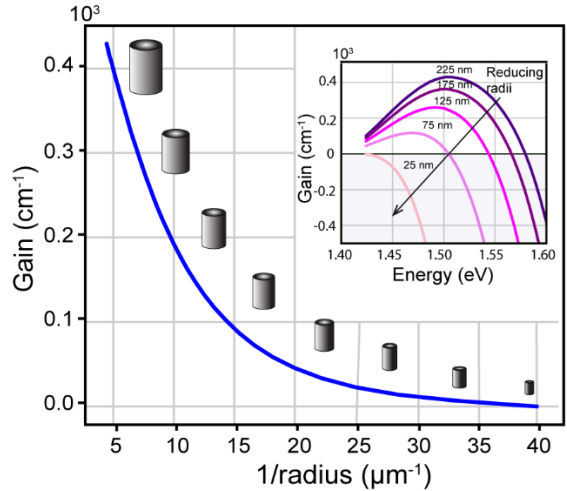

FIG. 6. Peak gain plotted against reciprocal of radius for decreasing GaAs nanocylinder radii from 225 nm to 25nm. Inset shows the corresponding gain spectrum for varying radii. Gain value as well as bandwidth decreases with shrinking nanostructure size implying higher threshold and limited cavity design space for lasing in smaller nanostructures

In conclusion, we analyzed the detrimental effects of inherent high surface recombination velocity of semiconductors on the gain threshold for lasing. Our analysis shows that p-doping and compressive strain incorporation provide remarkable improvement in static and dynamic properties of nanolaser. The differential gain, which is important for high speed modulation, is also improved in strained, p-doped GaAs structure. These improvements in threshold and differential gain are due to placing of both the quasi Fermi levels close to band edges. Our analysis shows that it is indeed possible to achieve lasing at room temperature with a suitable choice of semiconductor material parameters. We conclude that it is necessary for the active gain material for semiconductor nanolasers to be surface passivated, p-doped and strained in order to achieve room temperature lasing at lower fluence without thermally damaging the structure. We believe our work coupled with systematic studies of PL and lifetime in semiconductor nanostructures can provide important insights to develop semiconductor nanolasers and other all-dielectric metamaterial based active applications.

**SUPPLEMENTARY MATERIAL**

See Supplementary material for implementation details of FEM and atomistic numerical DFT simulations, physics and impact of strain on band structure and other band structure related parameters, impact of surface recombination on carrier density in structures comparable to carrier diffusion length and effects of heavy p-doping.

**ACKNOWLEDGMENTS**

This work was supported by the Science and Engineering Research Board (SERB) through Ramanujan fellowship (SB/S2/RJN-007/2017) and Early Career Research (ECR/2018/002452) grants. JT, SK and PK thank the Ministry of Human Resource Development (MHRD), Govt. of India for the research fellowship to undertake Ph.D. study.